\documentclass{template}

\usepackage{siunitx}
\usepackage{hyperref}
\usepackage{graphicx}%
\usepackage{hyperref}
\usepackage{subcaption}

\DeclareSIUnit\angstrom{\text{Å}}

\begin{document}

\makeatletter
\DeclareRobustCommand{\change}{%
  \@bsphack
  \leavevmode
  \color{red}%
  \@esphack
}
\DeclareRobustCommand{\stopchange}{%
  \@bsphack
  \normalcolor
  \@esphack
}
\makeatother

\pagestyle{fancy}
\rhead{\includegraphics[width=1.5cm]{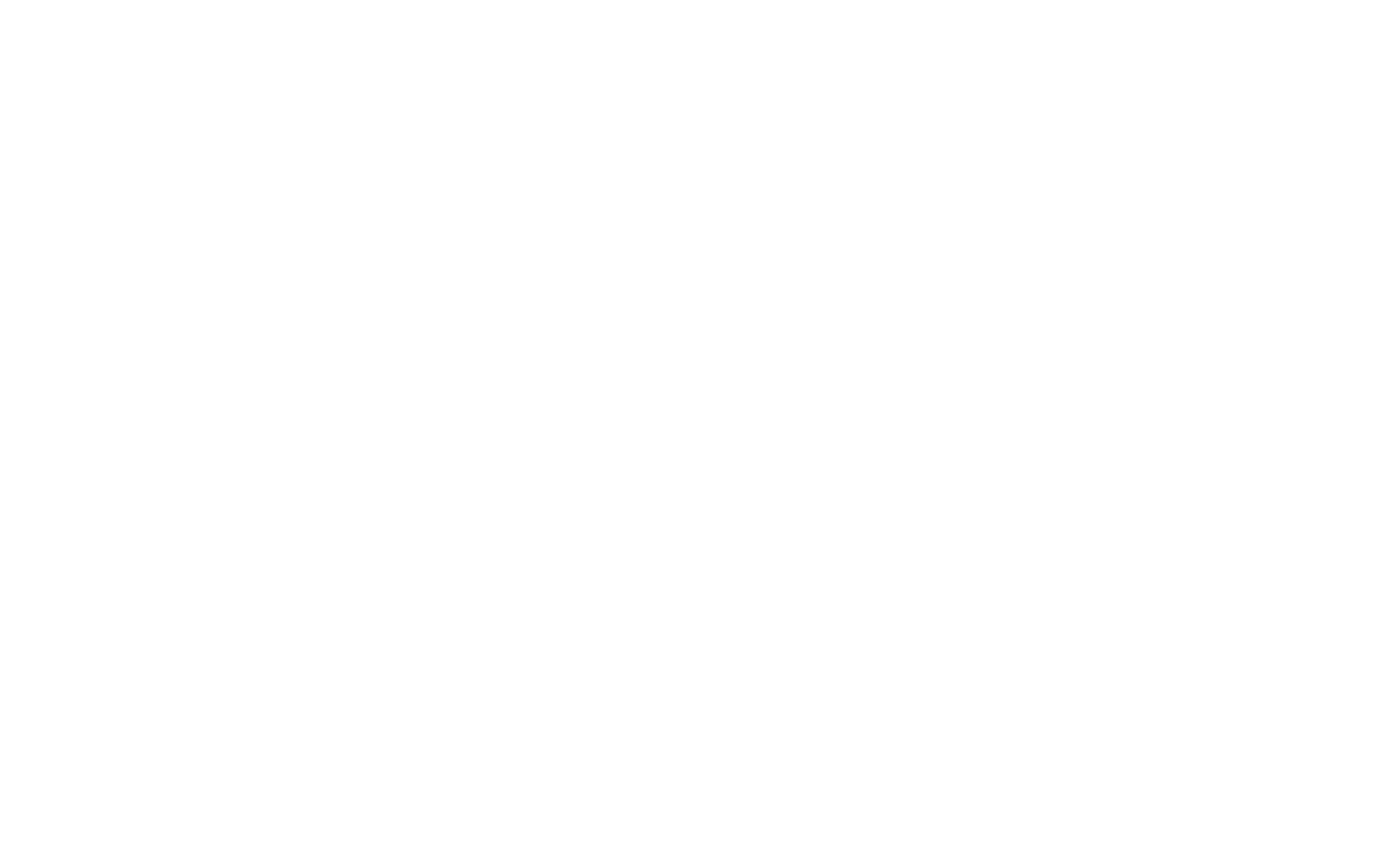}}

\title{Temperature-dependent anharmonic phonons in quantum paraelectric KTaO$_3$ by
first principles and machine-learned force fields}

\maketitle

\author{Luigi Ranalli}
\author{Carla Verdi}
\author{Lorenzo Monacelli}
\author{Georg Kresse}
\author{Matteo Calandra}
\author{Cesare Franchini*}

\begin{affiliations}
Luigi Ranalli\\
University of Vienna, Faculty of Physics and Center for Computational Materials Science, \\Kolingasse 14-16, 1090 Vienna, Austria\\
University of Vienna, Vienna Doctoral School in Physics, \\Boltzmanngasse 5, 1090 Vienna, Austria\\

Carla Verdi\\
University of Vienna, Faculty of Physics and Center for Computational Materials Science, \\Vienna, Austria\\

Lorenzo Monacelli\\
University of Rome, “Sapienza”, Dipartimento di Fisica,\\
Piaz.le Aldo Moro 5, 00185, Rome, Italy

Georg Kresse\\
University of Vienna, Faculty of Physics and Center for Computational Materials Science, \\Kolingasse 14-16, 1090 Vienna, Austria\\

Matteo Calandra\\
Department of Physics, University of Trento,\\ Via Sommarive 14, I-38123 Povo, Italy\\

Cesare Franchini\\
University of Vienna, Faculty of Physics and Center for Computational Materials Science, \\Kolingasse 14-16, 1090 Vienna, Austria\\
Alma Mater Studiorum - Universit\`a di Bologna,
Department of Physics and Astronomy "Augusto Righi",\\Bologna, 40127 Italy\\
Email Address: cesare.franchini@univie.ac.at

\end{affiliations}


\begin{abstract}
Understanding collective phenomena in quantum materials from first principles is a promising route toward engineering materials properties on demand and designing new functionalities. This work examines the quantum paraelectric state, an elusive state of matter characterized by the smooth saturation of the ferroelectric instability at low temperature due to quantum fluctuations associated with anharmonic phonon effects. 
The temperature-dependent evolution of the soft ferroelectric phonon mode in the quantum paraelectric KTaO$_3$ in the range $\si{0}-\SI{300}{K}$ is modelled by combining density functional theory (DFT) calculations with the stochastic self-consistent harmonic approximation assisted by an on-the-fly machine-learned force field.
The calculated data show that including anharmonic terms is essential to stabilize the spurious imaginary ferroelectric phonon predicted by DFT, in agreement with experiments. 
Augmenting the DFT workflow with machine-learned force fields allows for efficient stochastic sampling of the configurational space using large supercells in a broad and dense temperature range, inaccessible by conventional ab initio protocols.
This work proposes a robust computational workflow capable of accounting for collective behaviors involving different degrees of freedom and occurring at large time/length scales, paving the way for precise modeling and control of quantum effects in materials.  

\end{abstract}

\section{Introduction}

\begin{figure*}
    \begin{subfigure}{0.240\textwidth}
        \caption{}
        \includegraphics[width=\textwidth, trim={0 0 0 0},clip]{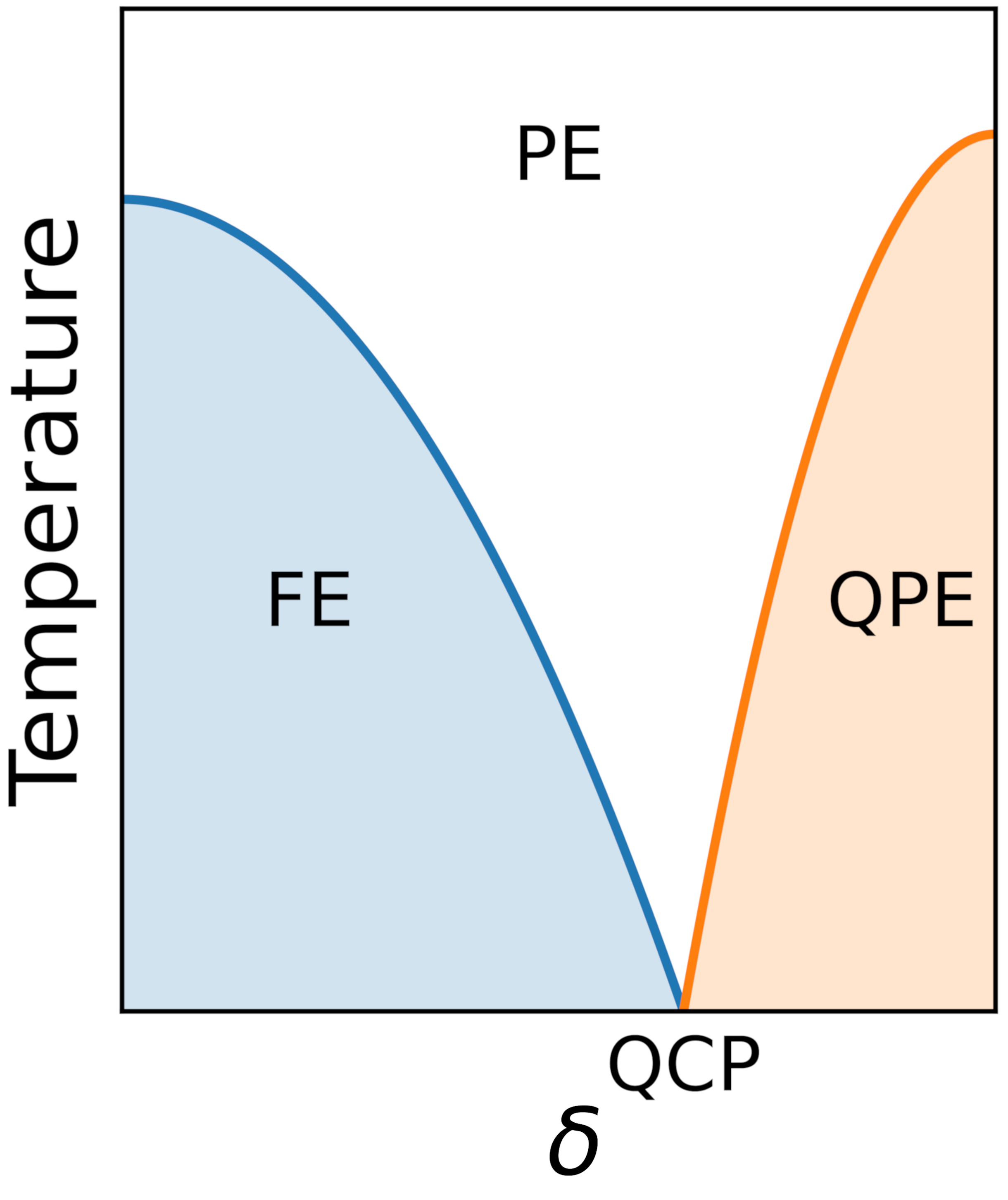}
        \label{fig:fig2_b}
    \end{subfigure}
    \hfill
    \begin{subfigure}{0.270\textwidth}
        \caption{}
        \includegraphics[width=\textwidth, trim={0 0 0 0},clip]{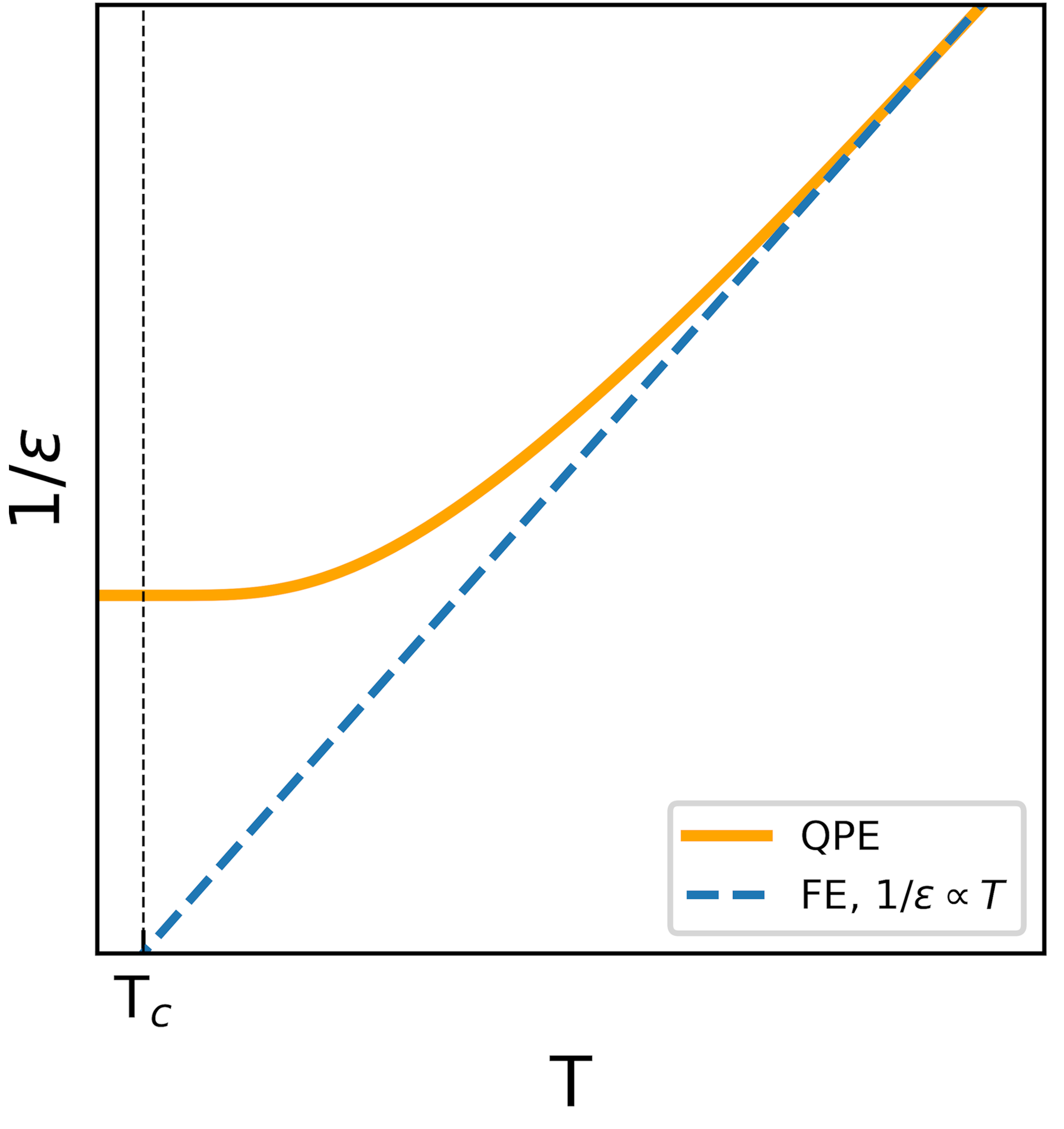}
        \label{fig:fig1_a}
    \end{subfigure}
    \hfill
        \begin{subfigure}{0.45\textwidth}
        \caption{}
        \includegraphics[width=\textwidth, trim={0 0 0 0},clip]{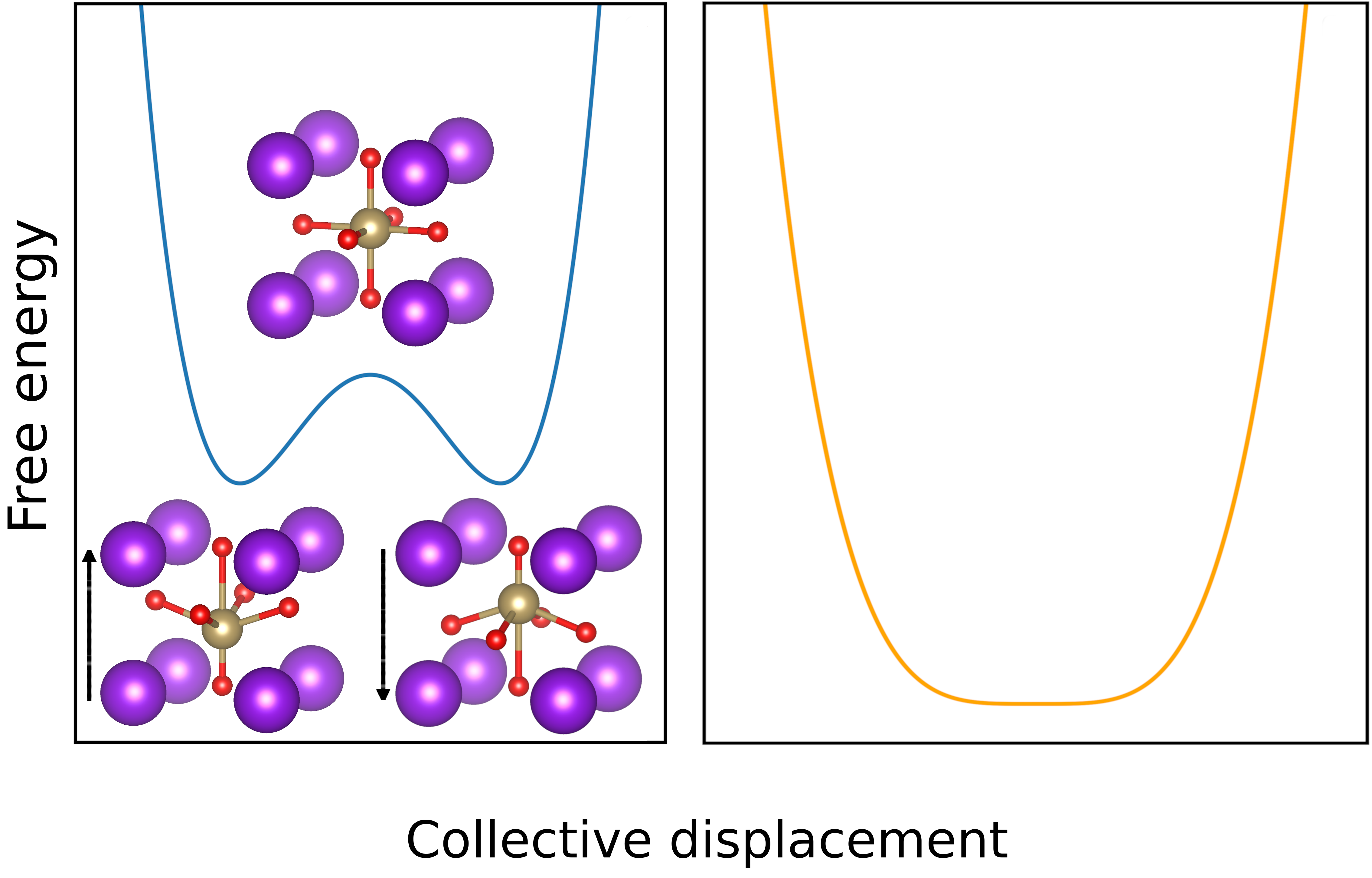}
        \label{fig:fig3_c}
    \end{subfigure}

  \caption{
  a) Schematic trend of the inverse dielectric constant as a function of temperature in the ferroelectric (FE) and quantum paraelectric (QPE) phase. In the FE phase 1/$\epsilon$ follows the characteristic linear behaviour (dashed blue line) leading to a divergence of $\epsilon$ at a critical temperature $T_c$ where the FE transition occurs. In a  quantum paraelectric (orange solid line), in the proximity of $T_c$ the onset of quantum  anharmonicity breaks down the linear regime establishing the QPE state.  
  b) Qualitative phase diagram of the crossover between the FE and QPE phases as a function of a quantum tuning parameter $\delta$; above a certain critical temperature, marked by the thick lines, the system lies in the paraelectric region. At $\SI{0}{K}$, variations of the quantum tuning parameter (due, for example, to strain, hydrostatic pressure or isotopic substitution) drive the FE-to-QPE transition across the so-called quantum critical point (QCP).
  c) Quantum free energy as a function of a collective atomic displacement in ferroelectric (left) and quantum paraelectric (right) systems. On the left, the quantum free energy as a function of the collective displacement amplitude induced by the unstable optical FE mode is here shown for a perovskite ABO$_3$ crystal. The soft FE phonon induces a FE displacement of the BO$_6$ octahedron (insets) which lowers the quantum free energy forming the typical double-well profile. In a QPE (right), the FE transition is inhibited due to quantum fluctuations and phonon anharmonicity.  
  The inclusion of the anharmonic self-energy contributions at 0~K in the SSCHA calculations flatten the internal energy removing the FE minima and establishing the QPE state.}
\end{figure*}

The quantum paraelectric phase is an eminent  example of a quantum state of matter. It is manifested by the suppression of the ferroelectric transition at low temperatures due to quantum fluctuations, as observed in strontium titanate (SrTiO$_3$, STO) and potassium tantalate (KTaO$_3$, KTO)~\cite{PhysRevB.19.3593}.
In these materials, the frequency of the polar transverse optical (TO) soft phonon mode 
does not turn unstable with decreasing temperature, as it occurs in regular ferroelectric materials such as BaTiO$_3$, but rather saturates and never reaches the zero frequency limit~\cite{Cowley1962,PhysRevLett.37.1155}. This behavior is associated with an unusual temperature dependence of the inverse dielectric constant at low temperature, at odds with the classical Curie-Weiss law~\cite{doi:10.1080/00150199508007850,Hideshi2016,Rowley2014a} (see \textbf{Figure~\ref{fig:fig2_b}}). 
Quantum paraelectric STO and KTO, laying on the border of the ferroelectric phase in the vicinity of the so-called quantum critical point (QCP)~\cite{sachdev_2000,Rowley2014a}, can be driven toward the ferroelectric region by applying minuscule external perturbations such as strain~\cite{PhysRevLett.104.227601}, hydrostatic pressure~\cite{PhysRev.151.378, PhysRevB.13.271}, or isotopic substitution~\cite{PhysRevLett.82.3540} (see Figure~\ref{fig:fig1_a}). For this reason, STO and KTO are also named incipient ferroelectrics.
How the transition occurs (i.e., how the long-wavelength \textbf{q}=0 ferroelectric TO phonon mode $\omega_{FE}$ becomes stable) is a complex matter challenging to decipher~\cite{Esswein}. However, there is a certain consensus on the importance of the anharmonic coupling between $\omega_{FE}$ and the low-T quantum lattice oscillations in determining the stabilization of $\omega_{FE}$ at helium temperatures in quantum paraelectrics~\cite{PhysRev.86.118,Cochran2006,Samara1973}. Within the Landau model of displacive phase transitions the quantum paraelectric (QPE) state can be viewed as the crossover between the ferroelectric (FE) state, characterized by the typical double-well free energy surface, and the standard parabolic paraelectric (PE) behaviour, resulting in quartic contributions~\cite{Lowndes1973} that flatten the free energy curve (Figure~\ref{fig:fig3_c}). In QPE materials, quantum fluctuations and anharmonic effects at low temperature stabilize the QPE solution that becomes the genuine (and highly perturbable) ground state of the system.

Abundant experimental data on the temperature decrease of the long-wavelength FE phonon and dielectric response in STO and KTO are available in literature~\cite{Perry,Rowley2014a}.
The measurements have been rationalized by various phenomenological models either based on extensions of the classic description of Slater~\cite{PhysRev.78.748}, such as the Barrett~\cite{PhysRev.86.118} and Vendik~\cite{Vendik} models~\cite{Hideshi2016}, or inspired by the Ginzburg-Landau-Wilson model such as 
the $\phi^4$-quantum field model~\cite{Rechester,Khmelnitskii}. From a computational modeling perspective, path-integral Monte Carlo 
calculations using model Hamiltonians have provided essential insights into the role of quantum fluctuations and non-linear response function on the suppression of the ferroelectric transition~\cite{PhysRevB.49.12596,PhysRevB.53.5047,PhysRevB.54.15714}.
Obtaining a microscopic description of the FE soft mode and the associated low-temperature non-linear behavior of the dielectric susceptibility is challenging.
Recently, valuable efforts to address this complex issue from first principles have been reported~\cite{PhysRevB.104.L060103}. 

In this work, we design an efficient protocol to compute the temperature-dependent frequency $\omega_{FE}(T)$ of the characteristic soft phonon mode and associated dielectric response from first principles with full quantum and anharmonic effects. We select KTO as case material since it retains the ideal cubic perovskite structure over a wide temperature range down to the paraelectric phase and is not subjected to possible complications associated with structural phase transitions (as is the case in STO). To include ionic quantum and thermal fluctuations we adopt the stochastic self-consistent harmonic approximation (SSCHA) method~\cite{doi:10.1080/14786440408520575,Monacelli2021}, integrated with an on-the fly machine learned force field (MLFF) scheme~\cite{Jinnouchi2019} for an efficient and accelerated exploration of the stochastic space.

Unlike alternative approaches such as extracting effective force constants from ab initio molecular dynamics (MD) 
trajectories~\cite{Hellman2011}, the SSCHA method is based on a rigorous variational method involving the quantum free energy functional that directly yields the anharmonic free energy from a suitably \\parametrized harmonic hamiltonian~\cite{Monacelli2021, Errea2013, Errea2014, Bianco2017, Monacelli2018}. 
To evaluate the necessary partial derivatives of the free energy of the auxiliary harmonic system the SSCHA method adopts a Monte Carlo (MC) stochastic procedure on 
a set of random ionic configurations generated in a chosen supercell following a Gaussian probability distribution. After the free energy functional
minimization, the renormalized phonon frequencies are obtained along with other quantities such as the
anharmonic phonon spectral functions and the anharmonic frequency linewidths.

The stochastic MC sampling is the most time-consuming part of the SSCHA workflow, as it requires the ground-state energy and interatomic forces for numerous supercell structures (of the order of a few thousands) obtained by randomly displacing the ions of the ideal crystal. To alleviate this huge computational cost a reweighting technique is introduced (the so-called importance sampling), that allows for the simultaneous execution of several minimization steps with the same ensemble, with the price of a statistical degradation. Despite this speed-up, the computational cost of reaching well-converged results easily becomes prohibitive for systems requiring large supercells, posing a limit for large-scale samplings.

Machine learning algorithms have proved to be capable of efficiently accelerating computer simulations, and represent an attractive way to cope with this technical limitation, enabling efficient speed-up and preserving accuracy~\cite{Schmidt2019,Csanyi2019,Unke}.
Here, we accelerate the evaluation of SSCHA stochastic averages using a MLFF. The force field potential is trained on the fly during MD runs at different temperatures and is then employed to feed the inputs required by the MC integrals, thus cutting down the computational cost of anharmonic phonons calculations by orders of magnitude.

\section{Results}

\subsection{Harmonic solution}

\begin{figure*}
    \begin{subfigure}{0.40 \textwidth}
        \caption{}
        \includegraphics[width=\textwidth, trim={0 0.5cm 0 0.35cm},clip]{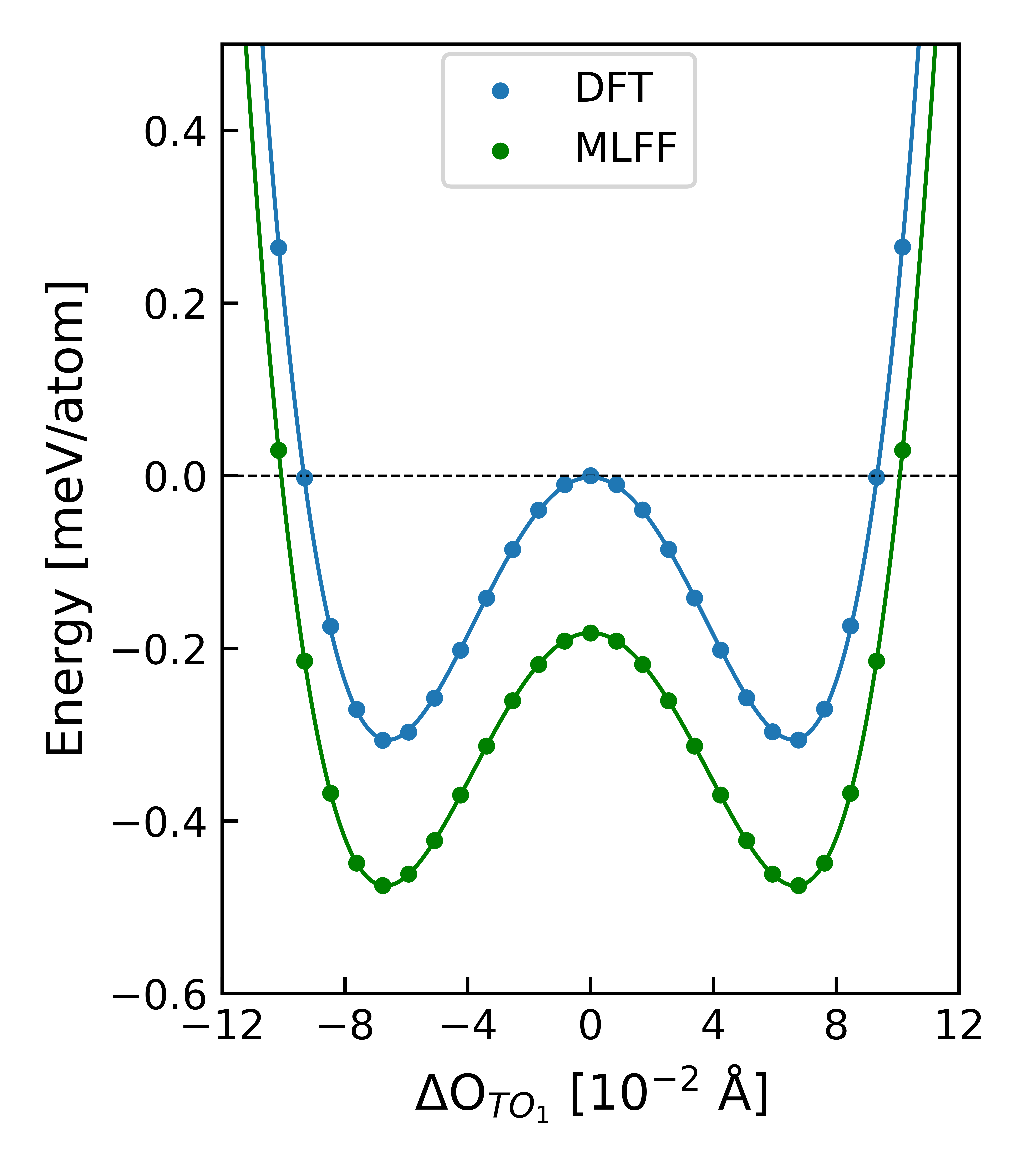}
        \label{fig:GS_a}
    \end{subfigure}
        \begin{subfigure}{0.5725\textwidth}
        \caption{}
        \includegraphics[width=\textwidth, trim={0 0.5cm 0 0.35cm},clip]{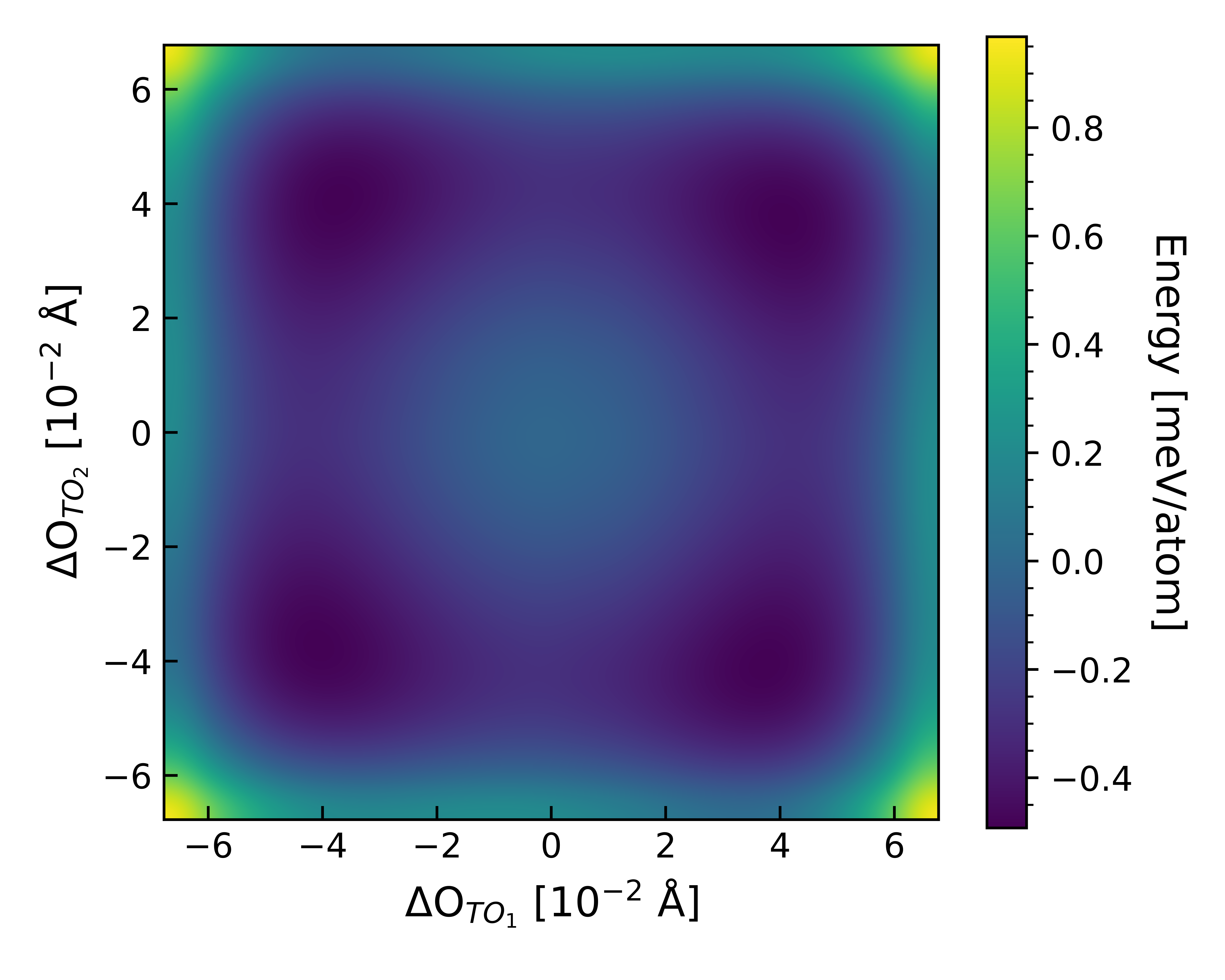}
        \label{fig:GS_b}
    \end{subfigure}
    \caption{Calculated ferroelectric instability. a) Electronic ground state energy as a function of the O displacement from the equilibrium position along the TO${_1}$ unstable soft mode. The saddle point in the DFT calculation corresponds to the ideal cubic structure and represents the reference zero. The MLFF prediction is in very good agreement with the ab initio results. b) MLFF prediction of the 2D energy landscape as a function of the displacements involved in both the TO${_1}$ and TO${_2}$ degenerate modes. Calculations performed on a 3$\times$3$\times$3 supercell.}
  \label{fig:pes}
\end{figure*}

    We begin by assessing the quality of our MLFF by inspecting the ground-state energy as a function of the FE displacement as compared to direct DFT calculations (details about the MLFF training are reported in Sec.~3).
    The results are shown in Figure~\ref{fig:GS_a}. 
    As expected, in the harmonic approximation DFT predicts the cubic structure to be a saddle point that is meta-stable against opposite polar displacements of the O and Ta ions within the TaO$_2$ plane, as depicted in the inset of Figure~\ref{fig:fig3_c}. This result is due to the neglect of anharmonic effects associated with the O-Ta-O polar vibration, as discussed in the next section. More precisely, this instability is controlled by two degenerate modes at $\Gamma$, TO${_1}$ and TO${_2}$: the TO${_1}$ soft mode represents a motion along the crystallographic $x$-axis, and it is completely analogous to the displacement along the $z$-axis described by TO$_2$.  
    
    The double well energy profile in Figure~\ref{fig:GS_a} shows that the displacive nature of the spurious ferroelectric phase transition in KTaO$_3$ is very well captured by MLFF both qualitatively and quantitatively. The DFT energy barrier amounts to $\SI{-0.306}{meV}$/atom at an O displacement of $\SI{0.059}{\text{Å}}$ in the TO${_1}$ direction, whereas the MLFF predicts $\SI{-0.293}{meV}$/atom at $\SI{0.068}{\text{Å}}$. 
    
    Since the TO${_1}$ and TO${_2}$ modes are degenerate, each linear combination of their associated eigenvectors is a legitimate solution of the problem for the non ferroelectric phase. This is shown in Figure~\ref{fig:GS_b}, which reports the 2D map of the ground-state energy due to the combined action of both FE modes as obtained from the MLFF. This 2D map indicates a 4-fold degenerate minimum at $\SI{-0.493}{meV}$/atom with respect to the saddle point located at (0,0).

\subsection{Anharmonicity}

Having tested the accuracy of the MLFF in accounting for the energy versus displacement profile, we now address its accuracy in obtaining the free energy Hessian with respect to the atomic positions, from which one can extract the anharmonic contributions to phonon frequencies. The occurrence of a ferroelectric transition as a function of temperature can be detected by tracing the eigenvalues of the free energy Hessian divided by the square root of the mass matrix (the so called positional free energy Hessian, see Equation 51 in Ref.~\cite{Monacelli2021}). An imaginary eigenvalue of the positional free energy diagnoses a ferroelectric transition. 

\begin{figure*}
    \begin{subfigure}{0.50\textwidth}
        \caption{}
        \includegraphics[width=\textwidth, trim={0 0 0 0},clip]{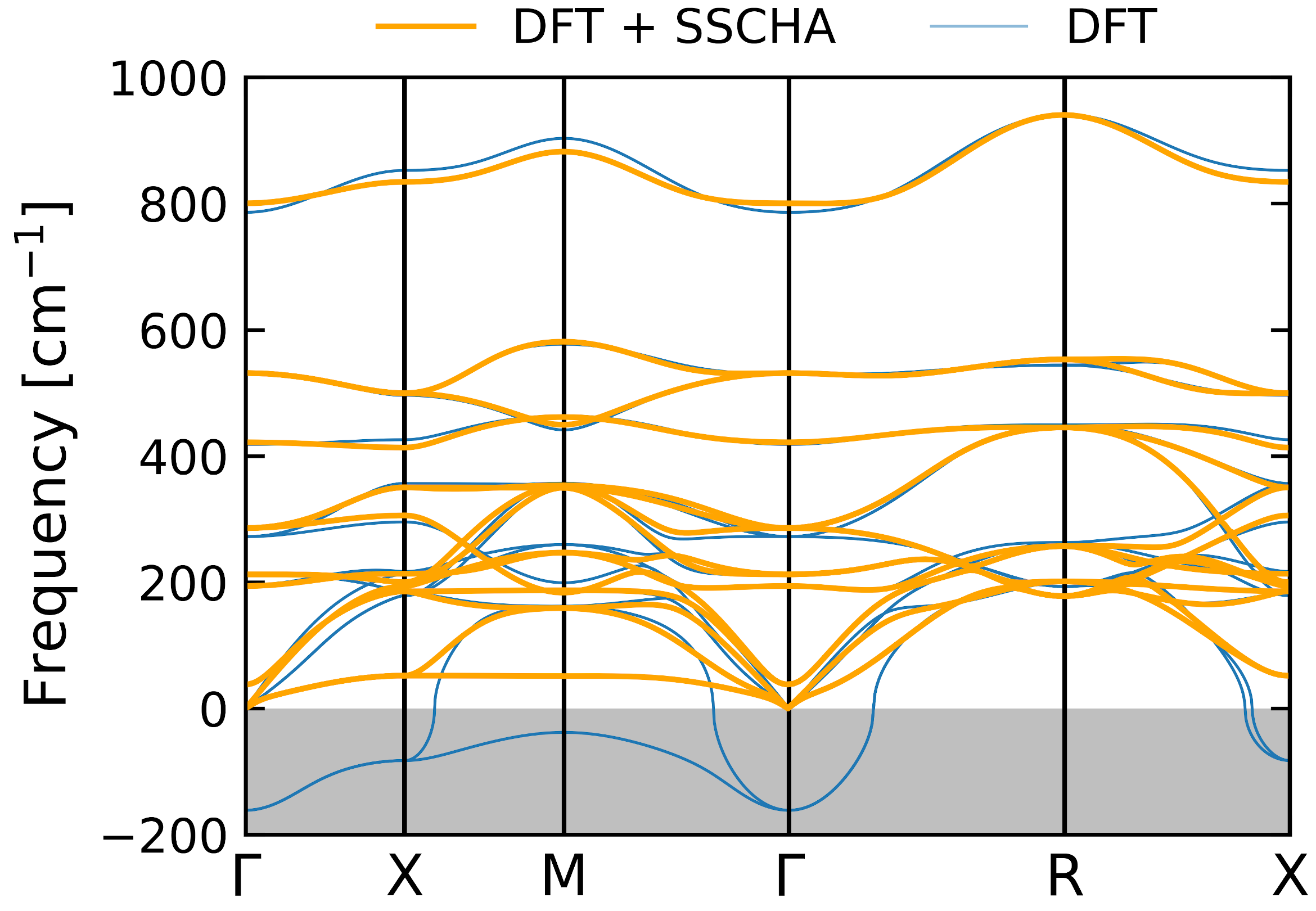}
        \label{fig:landscape_dispersion_a}
    \end{subfigure}
        \begin{subfigure}{0.4625\textwidth}
        \caption{}
        \includegraphics[width=\textwidth, trim={0 0 0 0},clip]{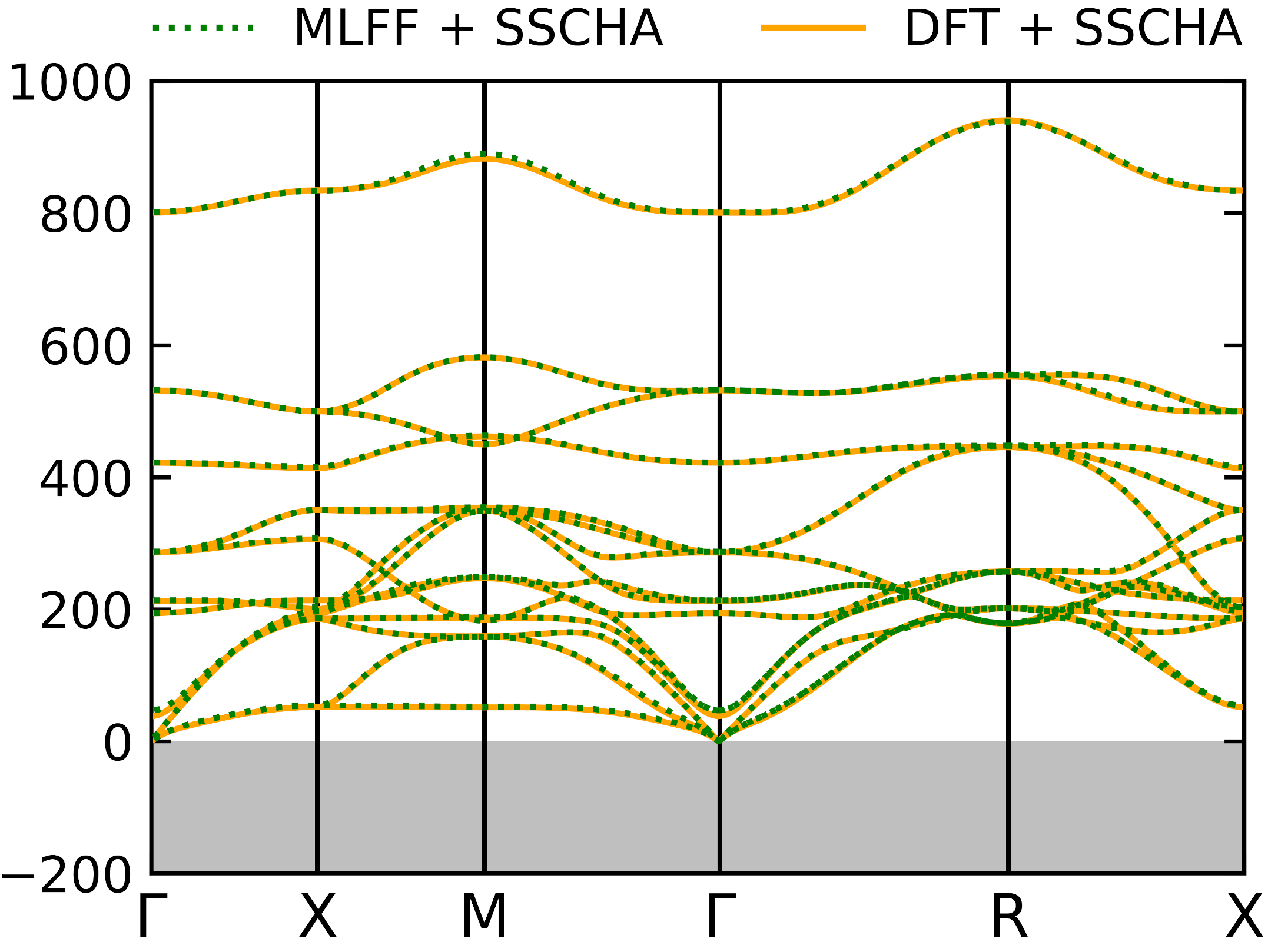}
        \label{fig:landscape_dispersion_b}
    \end{subfigure}
    \caption{Calculated \SI{0}{K} harmonic and anharmonic phonon dispersion. a) In DFT (blue lines), the harmonic phonon dispersion exhibits unstable FE TO$_1$ and TO$_2$ degenerate phonon modes at $\Gamma$. Within DFT+SSCHA (solid orange lines), the inclusion of the  bubble term (anharmonic effects) renormalizes the unstable modes, while keeping the other branches unaffected.
  b) Comparison between DFT+SSCHA and MLFF+SSCHA (green dotted lines) anharmonic phonon spectra. Calculations performed on a 2$\times$2$\times$2 supercell.}
  \label{fig:landscape_dispersion}
\end{figure*}

In Figure~\ref{fig:landscape_dispersion_a} we show the anharmonic phonon dispersions obtained from the eigenvalues of the positional free energy Hessian within the SSCHA at $T=0$ K by using DFT on a $2\times 2\times 2$ supercell and by including the self-energy correction in the free energy Hessian~\cite{Monacelli2021}. The phonon self energy is treated using the Bubble approximation that accounts for the most relevant three-phonon scattering processes. The result is compared with the DFT harmonic phonon dispersion. 
Not surprisingly, and in agreement with Figure~\ref{fig:GS_a}, the lowest energy optical phonon mode
(the degenerate ferroelectric modes, labeled TO$_1$ and TO$_2$) at $\Gamma$ is unstable in a large portion of the Brillouin zone,
in stark disagreement with experimental data~\cite{Perry}.
When quantum anharmonicity is switched on in the SSCHA method, the  $\SI{0}{K}$ FE instability is removed at DFT level. 
Remarkably, all other phonon modes are practically unaffected by the anharmonic correction. 
The calculated eigenvalue of the positional free energy for the TO$_1$ mode at $\Gamma$ is $\SI{38.11}{cm^{-1}}$. The contribution of the bubble term to the converged self-consistent dynamical matrix is $\SI{-40.26}{cm^{-1}}$. 

We now proceed to integrate on-the-fly MLFF with SSCHA in order to accelerate the calculation of the phonon renormalization and improve the stochastic sampling of the configurational space. The structure and technical details of the proposed protocol are described in Section~\ref{section:computational_methods}.
The results are shown in Figure~\ref{fig:landscape_dispersion_b}, which displays the comparison between the calculated DFT+SSCHA phonon dispersion along with the MLFF+SSCHA data. 
Note that for this comparison, here we employ an MLFF trained on a small $2\times 2\times 2$ supercell, i.e. the same supercell size adopted for the DFT+SSCHA calculations. For accurate finite-temeprature phonon properties we have employed a larger  $3 \times 3\times 3$ and a more robust MLFF as detailed in Section~\ref{section:computational_methods}.
The agreement is excellent in all regions of the phonon spectrum, including 
the TO$_1$ $\Gamma$-point frequency of $\SI{46.57}{cm^{-1}}$ and the $\SI{-37.15}{cm^{-1}}$ bubble correction, very close to the corresponding DFT+SSCHA estimate.
The neutron data at $\SI{20}{K}$ deliver a soft frequency at $\Gamma$ of $\SI{24.92}{cm^{-1}}$~\cite{Perry}, in line with our static dispersion data.

As the next level of complexity, we turn on finite temperature effects, which represent a key aspect to describe temperature-driven phase transitions and a huge obstacle for first principles approaches.
Investigating the soft mode temperature behavior using ab initio calculations in SSCHA is computationally prohibitive, due to the large number of supercells required to achieve well-converged results for many temperatures. To obtain accurate results, we adopt a $3\times3\times3$ supercell, with the corresponding MLFF trained to sample the configuration space up to 700~K.
This yields a higher frequency of $\SI{59.63}{cm^{-1}}$ for the bubble-term corrected soft mode when compared to the $\SI{46.57}{cm^{-1}}$ of the 2$\times$2$\times$2 case. Furthermore, the antagonistic bubble self-energy brings a lower correction of $\SI{-11.24}{cm^{-1}}$ to the SSCHA virtual phonons. We note that the bubble term correction on top of the converged dynamical matrix returns a negative shift of the soft mode of $\SI{15.49}{cm^{-1}}$ at $\SI{300}{K}$, hence its contribution is almost temperature independent.

\begin{figure*}
    \begin{subfigure}{0.47\textwidth}
        \caption{}
        \includegraphics[width=\textwidth, trim={0 0 0 0},clip]{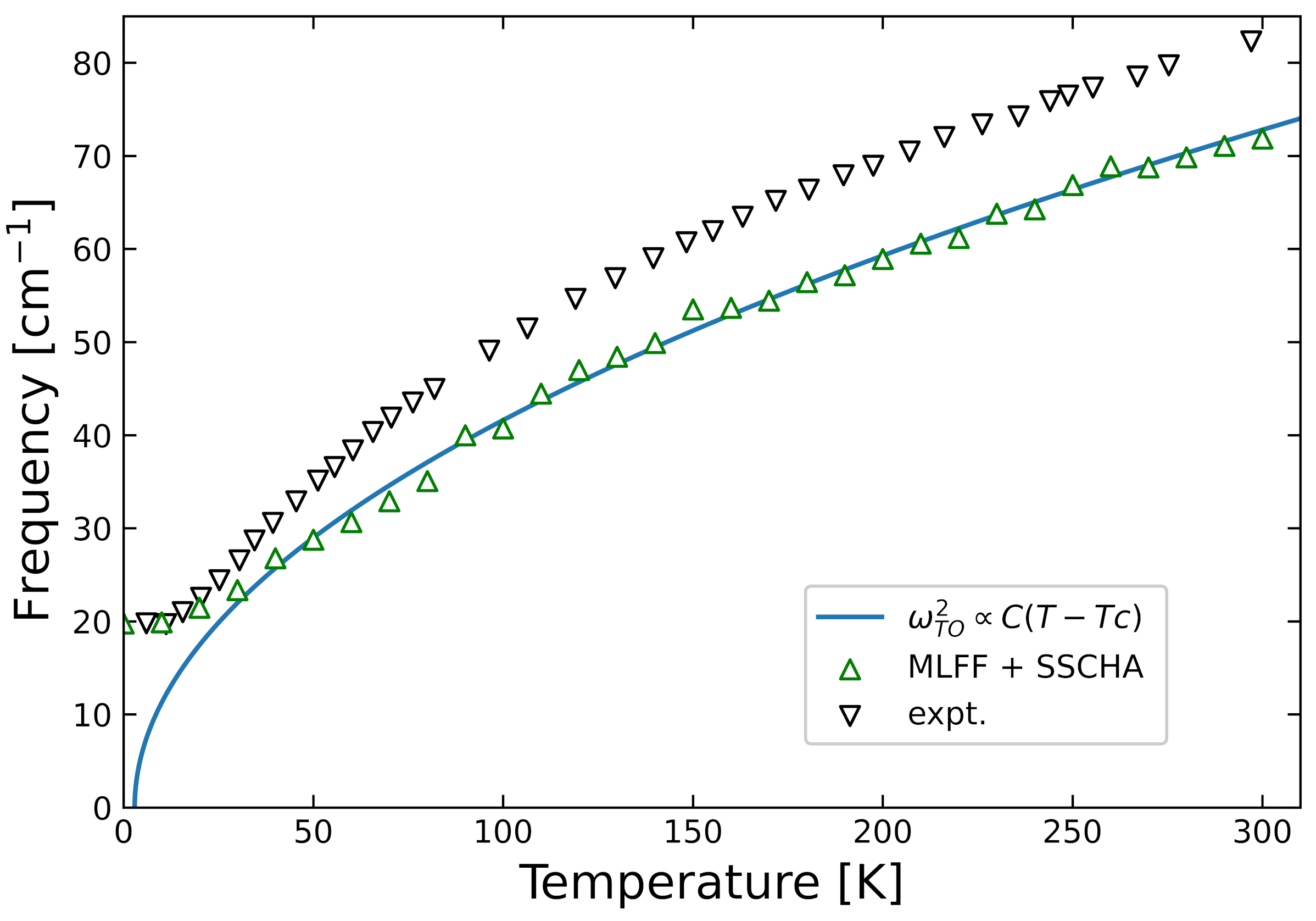}
        \label{fig:temperature}
    \end{subfigure}
        \hspace{0.54cm}
    \begin{subfigure}{0.48\textwidth}
        \caption{}
        \includegraphics[width=\textwidth, trim={0 0 0 0},clip]{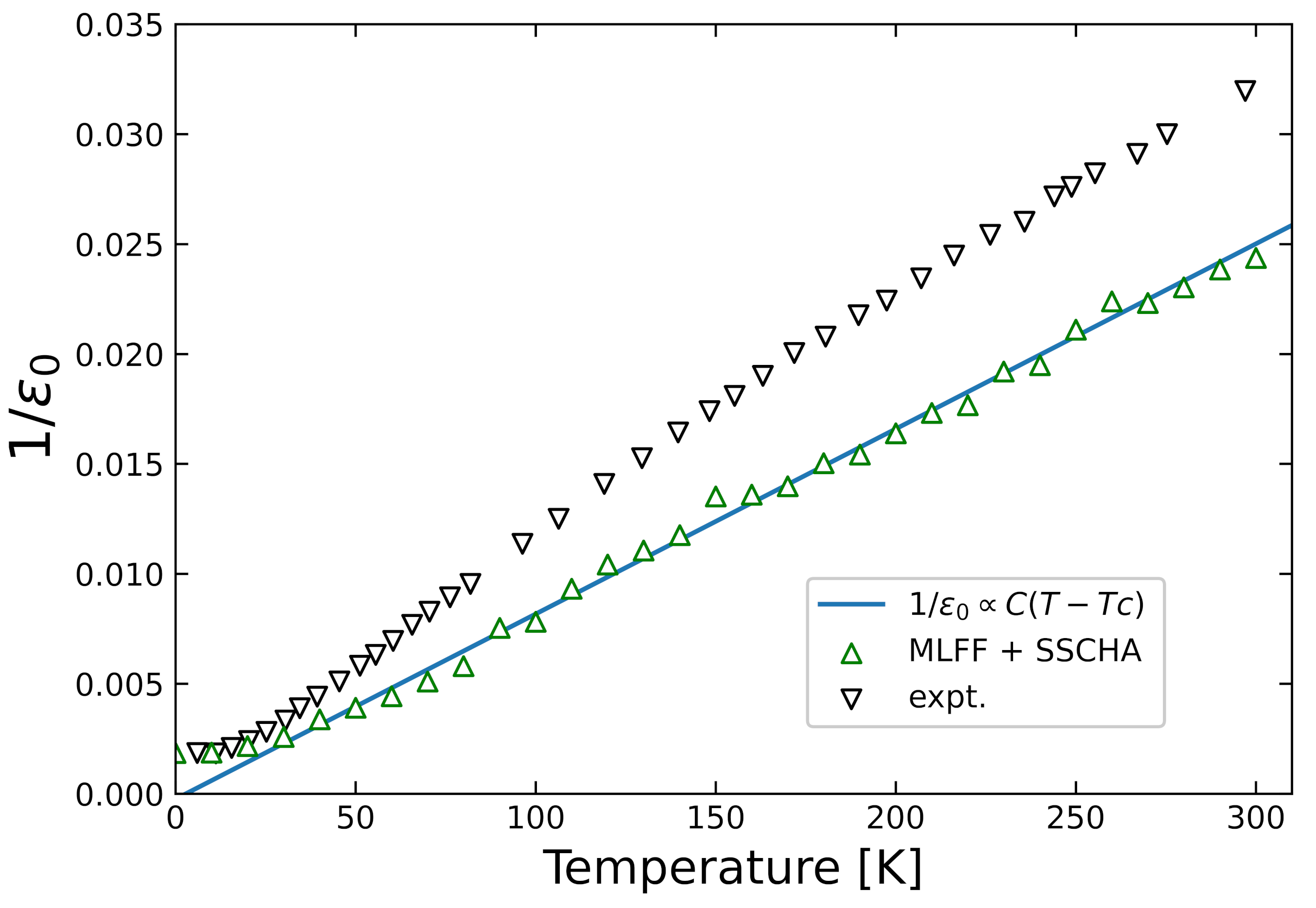}
        \label{fig:diel}
    \end{subfigure}
 \caption{Temperature evolution of the TO$_1$ $\Gamma$-point frequency (a) and of the 
 inverse of the dielectric constant (b). In (a), inelastic neutron scattering measurements~\cite{Vogt1995} (black empty triangles) and MLFF+SSCHA calculations with the inclusion of the bubble term  are shown. The computed frequencies are shifted down by $\SI{34.71}{cm^{-1}}$ in order to match the experimental low 
 temperature soft mode frequency, to better appreciate the character of the frequency plateaus. A square-root fit (blue line) shows the breakdown of the paraelectric regime  below $\SI{30}{K}$ and the onset of the quantum effects coupled with lattice anharmonicity, leading to a frequency plateau near $\SI{0}{K}$. Analogously, the blue line in (b) shows the expected classical Curie-Weiss behaviour (see text) of the dielectric constant, calculated from the shifted frequencies of (a). Calculations performed on a 3$\times$3$\times$3 supercell.}
 \label{fig:temperature_dependence}
\end{figure*}

The soft mode temperature behavior predicted using the MLFF is plotted in Figure~\ref{fig:temperature}, aligned with the low-T experimental data taken from Ref.~\cite{Vogt1995}.
The MLFF predictions reproduce fairly well the measured trend demonstrating how the efficient integration of electronic structure methods, stochastic approaches and artificial intelligence schemes provides a powerful computational framework that gives us the tools to explore previously inaccessible regimes and behaviors.

In particular, our data correctly predict the plateau below $\SI{30}{K}$, in line with the expected theoretical trend collected in Figure~\ref{fig:fig2_b}, proving the ability to capture the crucial role played by anharmonic effects in setting up the  the quantum paraelectric state. 
By assuming the Lydanne-Sachs-Teller (LST) relation~\cite{PhysRev.59.673} 
$\frac{\omega^2_\textup{LO}}{\omega^2_\textup{TO}} = \frac{\epsilon_0}{\epsilon_\infty}$
connecting the optical phonon frequencies to the static and high frequency (above the phonon frequencies, but below any electronic energy scale) dielectric constants, and considering only the temperature dependence of the dielectric constant due to the TO dependence, 
we plot in Figure~\ref{fig:diel} the inverse of the dielectric constant $1/\epsilon_0$ as a function of temperature, together with the one extracted from the experimental data. The computed values of 4.635 and $\SI{212.68}{cm^{-1}}$ were adopted, respectively, for $\epsilon_\infty$ and the longitudinal optical mode $\omega_\textup{LO}$. 
A square-root fit of the frequency data, or equivalently a linear fit of the temperature-dependent inverse dielectric constant, clearly shows how the classic regime breaks down below $\SI{30}{K}$, where the quantum nuclear motion coupled with anharmonicity starts prevailing.

\begin{figure*}
    \begin{subfigure}{0.43\textwidth}
        \caption{}
        \includegraphics[width=\textwidth, trim={0 0 0 0},clip]{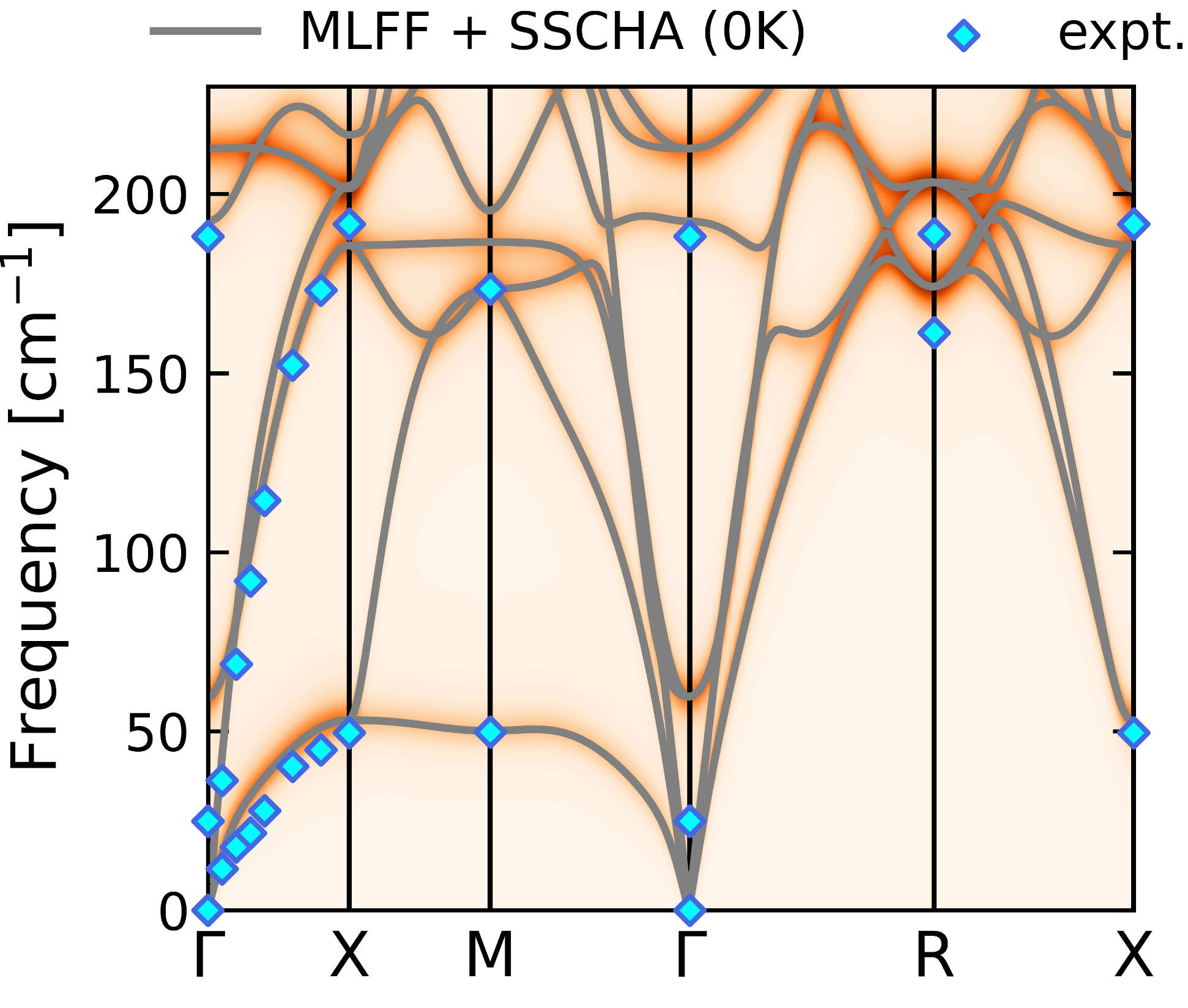}
        \label{fig:figure5_a}
    \end{subfigure}
        \hspace{0.53cm}
    \begin{subfigure}{0.48\textwidth}
        \caption{}
        \includegraphics[width=\textwidth, trim={0 0 0 0},clip]{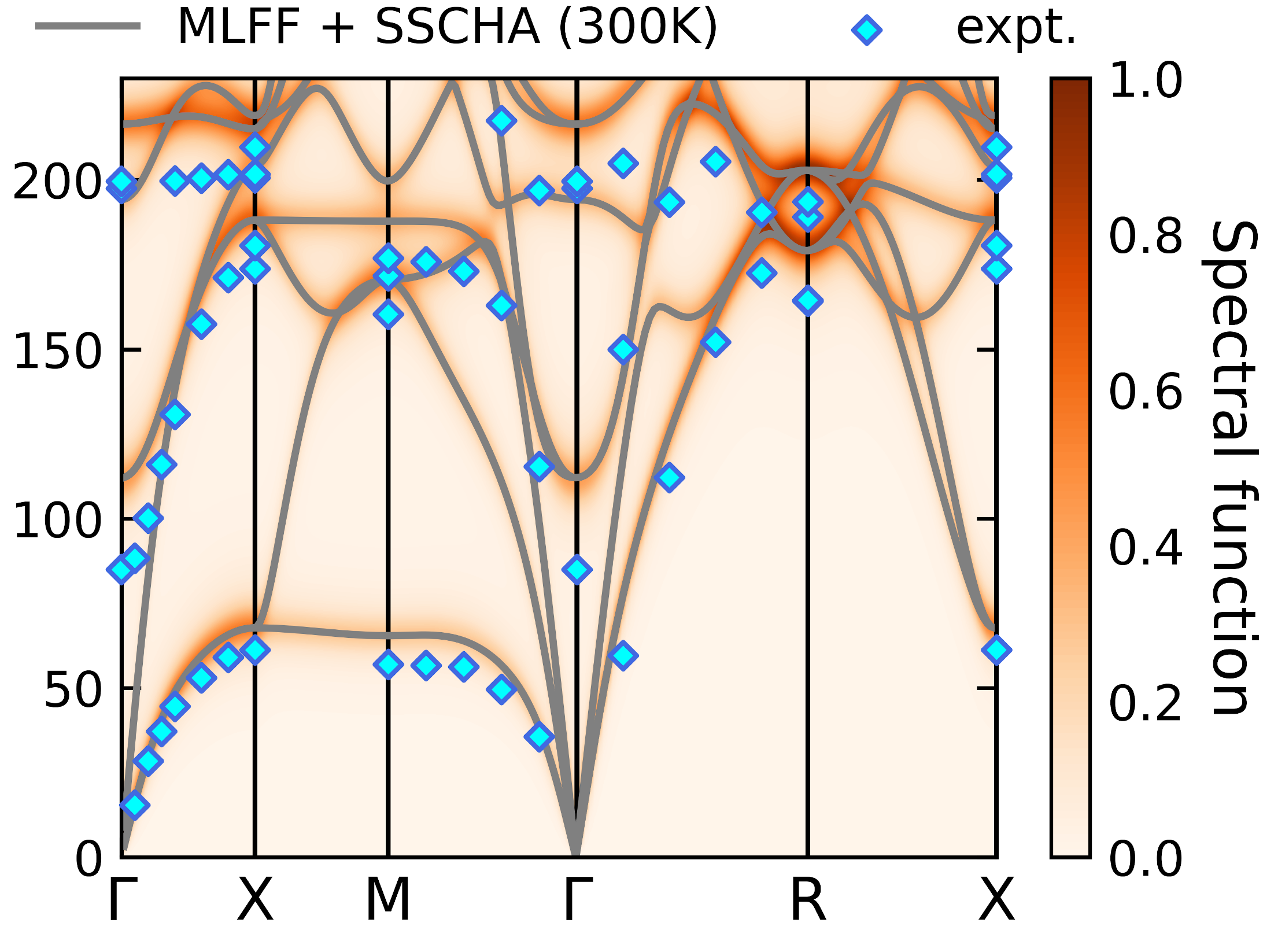}
        \label{fig:figure5_b}
    \end{subfigure}
 \caption{Color plot of the normalized phonon spectral function along the high symmetry path at (a) $\SI{0}{K}$ and (b) $\SI{300}{K}$. In both cases the maxima of the spectral function in darker red coincide with the static dispersion (full lines). The results are compared with the experimental data at $\SI{20}{K}$ and $\SI{296}{K}$~\protect\cite{Perry} (diamonds). Calculations performed on a 3$\times$3$\times$3 supercell.}
\end{figure*}

We conclude by computing the phonon spectral function, which accounts for dynamical effects not included in the phonon spectra shown Figure~\ref{fig:landscape_dispersion} and whose peaks give the phonon quasiparticles measured in inelastic neutron scattering.
In~\ref{fig:figure5_a} and Figure~\ref{fig:figure5_b} the phonon (bosonic) spectral function at $\SI{0}{K}$ and $\SI{300}{K}$ on a $3\times3\times3$ supercell is plotted on the high-symmetry $k$-path  and compared with available experimental data~\cite{Perry} and corresponding static dispersions.  We note that 
 the maxima of the spectral function coincide with the static dispersions (relative to the same    3$\times$3$\times$3 supercell) at bot selected temperatures. In this case the dynamical Bubble correction does not offer any substantial change on the static one, such as the appearence of satellite excitations. The agreement with the measured data is very good, even though the soft TO mode is overestimated by a few tens of cm$^{-1}$ as pointed out above.

\section{Computational methods}
\label{section:computational_methods}
\begin{figure}[h]
  \includegraphics[width=1\linewidth]{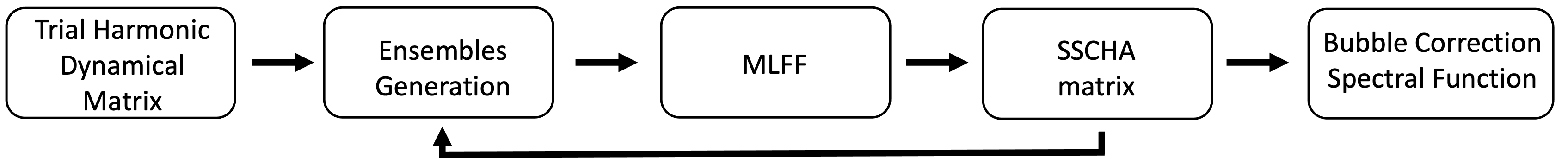}
  \caption{Sketch of the MLFF+SSCHA workflow. A first guess of the harmonic dynamical matrix is calculated at the DFT level or using the MLFF and employed by SSCHA in the generation of the first population (ensembles generation). The energies and forces for the displaced ensemble are computed employing the MLFF, and the gradient of the free energy is minimized in SSCHA (SSCHA matrix). The obtained SSCHA dynamical matrices are subsequently employed for the generation of the second population and the loop continues until convergence (arrow). After reaching convergence the bubble correction at both static and dynamical level can be calculated.}
  \label{fig:workflow}
  \end{figure}

The calculations were performed integrating the Vienna \emph{Ab initio} Simulation Package (VASP)~\cite{Kresse1993, Kresse1996} with the stochastic self-consistent harmonic approximation (SSCHA) method~\cite{doi:10.1080/14786440408520575,Monacelli2021}
and the phonon package Phonopy~\cite{phonopy}. 
All DFT calculations were executed at the meta-GGA level using the Strongly Constrained and Appropriately Normed (SCAN) functional~\cite{Sun2015} and projector augmented wave (PAW) potentials~\cite{Kresse1999}, with a plane-wave cutoff of \SI{800}{eV}.
For this purpose, a computational workflow was constructed, including a VASP-to-SSCHA interface (link provided in "Supporting Information") and the use of a MLFF in computing energies and forces associated with the generated ensembles within the SSCHA framework.
The workflow is schematically depicted in Figure~\ref{fig:workflow} and consists of the following parts:

\begin{enumerate}
    \item Construction of the \textit{trial harmonic dynamical matrix} built on a cubic unit cell of KTaO$_3$ at the experimental lattice constant of $\SI{3.9842}{\angstrom}$~\cite{Samara1973}.
    \item \textit{Ensembles generation} and self-consistent SSCHA loop. The free energy minimization was executed through a force constant gradient descent, keeping the atomic positions and volume fixed. Around 15000 ensembles were needed in the self-consistency loop at 0~K (the last 5000 ensembles were also employed at the end of it for computing the bubble correction).
    \item Speed-up of the stochastic MC sampling by means of the MLFF.
    This is the core part of the methodology.
    The MLFF is trained on the fly through MD calculations following the procedure outlined in Ref~\cite{Jinnouchi2019}.
    Two MLFF datasets were built: (i) A light one on a 2$\times$2$\times$2 supercell with a 3$\times$3$\times$3 
    $\Gamma$-centered $k$-point mesh, and (ii) a second one on a 3$\times$3$\times$3 supercell with a 2$\times$2$\times$2 Monkhorst-Pack $k$-point mesh.
    The MD runs were executed at increasing temperatures, sampling 50000 steps at each temperature with a time step of \SI{2}{fs}, and using the 
    final structure at each temperature as the starting configuration for the higher ones. 
    A Langevin thermostat was employed~\cite{Hoover1982}, with a friction coefficient of $\SI{10}{ps^{-1}}$ for each atomic species. For the descriptors representing the local atomic environments in the MLFF~\cite{Jinnouchi2020}, the cut-off radius for the two- and three-body atomic density distributions was set to \SI{6}{\angstrom}, and the Gaussian broadening to \SI{0.3}{\angstrom}. The same weight was assigned to both descriptors. 

The $2\times2\times2$ supercell dataset was used to train an MLFF to assess the accuracy of MLFF+SSCHA against direct DFT+SSCHA results obtained at $\SI{0}{K}$ on the same supercelll size (Figure~\ref{fig:landscape_dispersion_b}). For this training 225 structures were authomatically extracted running 3 MD simulations at \SI{1}{K}, \SI{100}{K} and \SI{200}{K}.
The resulting root mean square errors (RMSE) for energies and forces for 100 random ensembles generated at \SI{0}{K} are \SI{0.16}{meV/atom} and \SI{0.020}{eV/\text{Å}}, respectively.

To achieve converged and accurate results at higher temperatures, we trained the second MLFF by sampling configurations up to 700~K in a larger $3\times3\times3$ supercell.
This is required in order to precisely compute the energies and forces related to the Gaussian distributed SSCHA ensembles up to room temperature.
This dataset contains 680 reference configurations sampled by MD runs between \SI{1}{K} and \SI{700}{K}, with steps of \SI{100}{K}. A direct validation of MLFF+SSCHA was in this case not possible, since performing DFT+SSCHA on a 3$\times$3$\times$3 supercell turned out to be computationally prohibitive. To validate this MLFF, 
we computed the 1D and 2D energy landscape displayed in Figures~\ref{fig:GS_a} and~\ref{fig:GS_b}, showing excellent agreement with the DFT data. 
With this model we obtained RMSEs very close to the ones obtained for the first ($2\times2\times2$) MLFF, \SI{0.16}{meV/atom} and \SI{0.027}{eV/\text{Å}}, for energies and forces respectively, indicating a similar accuracy of both models.  
To guarantee a good performance of the MLFF up to \SI{300}{K}, 100 ensembles were also generated at this temperature, and the associated RMSEs are \SI{0.28}{meV/atom} and \SI{0.056}{eV/\text{Å}}, slightly larger than those associated to \SI{0}{K} calculations.

\item \textit{Bubble correction and spectral function}. 
The converged SSCHA matrix is finally corrected by the bubble self-energy term involving a three-phonons vertex, within the framework of a static Green's function approach where the free propagator is the one associated to the SSCHA Hamiltonian. The quartic SSCHA correction required to recover the full free energy curvature beyond the bubble term brought a negligible contribution to the soft mode at $\SI{0}{K}$ and $\SI{300}{K}$ for both the 2$\times$2$\times$2 and 3$\times$3$\times$3 supercells. Therefore, one can conclude that the bubble term alone encodes virtually all relevant anharmonic effects taking place in the real system up to $\SI{300}{K}$. To conduct a more robust comparison with measured phonon properties we have computed the spectral function within a dynamical treatment that removes the conservation of energy and momentum in the multi-phonon processes.~\cite{Monacelli2021} 
An interpolation on a 13$\times$13$\times$13 k-mesh was employed in order to converge the peak positions.

\end{enumerate} 

\section{Conclusions}

In this work we have illustrated that the inclusion of machine-learned force fields in the 
stochastic self-consistent harmonic approximation
allows the account of finite-temperature phonon properties of quantum paraeletric materials, a goal inaccessible by standard first-principles approaches. The application of this novel computational scheme to the incipient ferroelectric KTaO$_3$ provides an excellent prediction of high-order anharmonic effects as a function of temperature, a key ingredient to characterize the unusual low-T blocking of ferroelectric instabilites due to quantum and anharmonic fluctuations. 
In particular, the peculiar temperature-dependent frequency plateau near \SI{0}{K} is correctly reproduced, thus catching the essential quantum paraelectric nature of KTaO$_3$. 

The proposed ML-assisted method is capable to reproduce the $\Gamma$-point TO soft mode renormalization at \SI{0}{K} with only 225 MD ab initio calculations, a small fraction of the thousands of calculations required in the standard SSCHA method.
Adopting a larger MLFF training dataset of 680 structures sampled up to \SI{700}{K}, an accurete prediction of finite-temeprature phonon properties was computed between $\SI{0}{K}$ and $\SI{300}{K}$. Each temperature point would require several thousands of conventional ab-initio calculations, an overwhelming computational cost when compared to the few hundreds needed to build an accurate MLFF. 

Although MLFF+SSCHA yields a convincing qualitative trend of finite-temperature properties, a quantitative account of the strongly temperature dependent soft mode is still out of reach, including the recently reported minuscule upturn of the dielectric constant just above the absolute zero~\cite{Rowley2014a}.
This would probably require the adoption of a more sophisticated exchange-correlation functional (beyond meta-GGA) and a careful treatment of (finite-temperature) volume effects: both these aspects will be the focus of future research. 

Overall, the SSCHA+MLFF approach offers a convenient and flexible methodology for accessing complex anharmonicities as a function of temperature in materials at large system size. 
The method will enable to predict and interpret finite-temperature quantum effects expanding the predictive power of computational modelling and providing essential support to experiment.

\medskip
\textbf{Supporting Information} \par 
The "vasp-phonopy-sscha" interface is publicly available as a github repository~\url{https://github.com/QuantumMaterialsModelling/vasp-phonopy-sscha}.

\medskip
\textbf{Acknowledgements} \par 
This work was supported by the Austrian Science Fund (FWF) projects I 4506 (FWO-FWF joint project) and SFB TACO (F81). The computational results presented have been achieved in part using the Vienna Scientific Cluster (VSC).

\medskip

\bibliographystyle{MSP}
\bibliography{bibliography.bib}

\end{document}